\documentclass[
superscriptaddress, notitlepage,
nofootinbib,
 amsmath,amssymb,
 aps, prd,
]{revtex4-2}

\usepackage{graphicx}
\usepackage{dcolumn}
\usepackage{bm}
\usepackage{hyperref}
\usepackage{slashed}
\usepackage[dvipsnames]{xcolor}
\usepackage{float}

\begin{document}


\title{Diffractive rho + lepton pair production  at an electron-ion collider}

\author{W.~Cosyn}
\email[ E-mail: ]{wcosyn@fiu.edu}
\affiliation{%
 Department of Physics, Florida International University,  Miami, Florida 3199, USA
}%
\affiliation{
 Department of Physics and Astronomy, Ghent University, B9000 Ghent, Belgium
}

\author{ B.~Pire}
\email[ E-mail: ]{bernard.pire@polytechnique.edu}
\affiliation{ Centre de Physique Th\'eorique, CNRS, \'Ecole Polytechnique, I. P. Paris,
  91128 Palaiseau,     France  }

\date{\today}

\begin{abstract}
In high energy electron-ion colliders, a new way to probe nucleon structure becomes available through diffractive reactions, where the incident particle produces a very energetic almost forward particle. QCD describes these reactions as due to the exchange of a Pomeron which may be perturbatively  described as a  dressed two-gluon state, provided a hard scale allows the factorization of the amplitude in terms of two impact factors convoluted with a Pomeron propagator. We consider here a process where such a description allows to access hadronic structure in terms of the generalized parton distributions, namely the electroproduction of a forward $\rho$ meson and a timelike deeply virtual photon, separated by a large rapidity gap.  We explore the dependence of the cross section on the kinematic variables and study the dependence on the non-perturbative inputs (generalized parton distributions, distribution amplitude).  Our leading order studies show the cross section is mainly sensitive to the GPD model input, but the small size of the cross sections could prohibit straightforward analysis of this process at planned facilities.
\end{abstract}

\maketitle


\section{Introduction}
The advent of high luminosity  high energy electron-ion colliders \cite{Boer:2011fh,Accardi:2012qut,AbdulKhalek:2021gbh,AbelleiraFernandez:2012cc,Anderle:2021wcy} will allow to open a new chapter in the quest for the understanding of quark and gluon confinement in hadrons, through a precise tomography of the nucleon enabled - among  various tools - by the extraction of quark and gluon generalized parton distributions (GPDs)\cite{Diehl:2003ny,Belitsky:2005qn}. Besides the famous exclusive processes which have been studied in the last 20 years - namely deep electroproduction of a photon (deeply virtual Compton scattering, DVCS) or a meson (deeply virtual meson production, DVMP) and their timelike related processes - a new class of processes is worth studying which adds the merits of diffractive processes. These diffractive processes have been shown to constitute a sizeable part of the total cross section at very high energy.

\begin{figure} [htb]
    \centering
    \includegraphics[width=.4\textwidth]{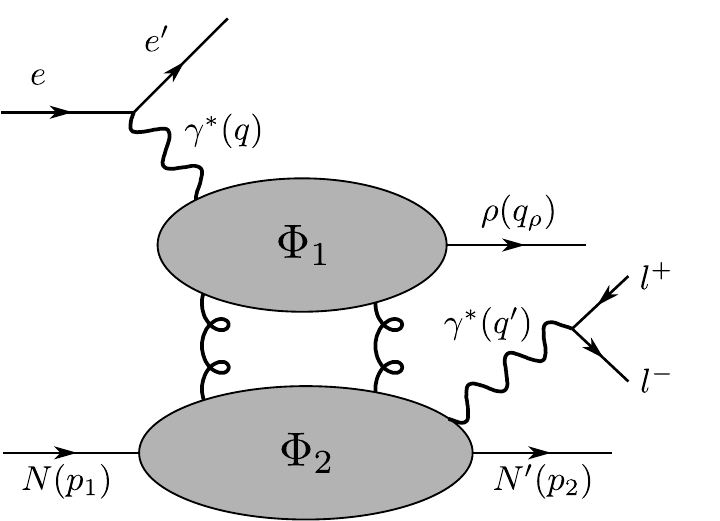} 
    \caption{The diffractive $\rho$ + lepton pair amplitude is written in the  $k_T$ -factorization approach as the convolution of two impact factors $\Phi_1$ and $\Phi_2$ and the Pomeron propagator, which at lowest order is a two gluon exchange.  Labels between brackets are the particle four momenta.
    }
    \label{fig:process}
\end{figure}

In a hard regime characterized by a hard scale $Q$, a diffractive reaction is seen as the scattering of a small transverse-size  [$O(\frac{1}{Q})$] colorless dipole on a nuclear target. This justifies the use of perturbative QCD methods for the description of the process. In the  Regge inspired $k_T$-factorization approach which is known to be applicable at high energy, total invariant mass $W \gg Q \gg \Lambda_{QCD}$, one writes the scattering amplitude in terms of two impact factors  with, at leading order, a two ``Reggeized'' gluon exchange in the $t$-channel. The Born amplitude may be calculated using a two gluon exchange (see diagram of Fig.~\ref{fig:process}), while higher order QCD corrections are taken into account by applying  Balitsky-Fadin-Kuraev-Lipatov (BFKL) evolution techniques~\cite{Fadin:1975cb, Kuraev:1977fs, Balitsky:1978ic}. 

In a previous paper \cite{Pire:2019hos}, it has thus been proposed to replace in the timelike Compton scattering (TCS) reaction \cite{Berger:2001xd}
\begin{equation}
\gamma(r, \varepsilon) + N(p_1,\lambda_1) \rightarrow  \gamma^*(q', \varepsilon') +  N'(p_2,\lambda_2)\,,
\label{TCS}
\end{equation}
the incoming photon by a Pomeron ($\mathcal{P}$) 
\begin{equation}
\mathcal{P}(r) + N(p_1,\lambda_1) \rightarrow  \gamma^*(q', \varepsilon') +  N'(p_2,\lambda_2)\,.
\label{PTCS}
\end{equation}
Here, $\varepsilon', \varepsilon$ denote the polarization states of the photons, while $\lambda_{1},\lambda_2$ denote those of the nucleon.  Momentum labels follow those of Fig.~\ref{fig:process}.
As Fig.~\ref{fig:process} shows, this subprocess (\ref{PTCS}) may be extracted from  the study of the  diffractive process
\begin{equation}
\gamma^*(q, \varepsilon) + N(p_1,\lambda_1) \rightarrow \rho^0(q_\rho,\varepsilon_\rho) + \gamma^*(q', \varepsilon') +  N'(p_2,\lambda_2)
\label{processGeneral}
\end{equation}
in the adequate kinematical domain (see details below) where  the overall process is calculated in the $k_T$ -factorization approach, and $r\equiv q-q_\rho$. This applies collinear QCD factorization to describe i) the Pomeron production by an impact factor where the $\rho$ meson enters by its distribution amplitude (DA) and ii) the Pomeron-nucleon interaction, with the hadronic response parametrized by non-perturbative GPDs, as depicted further down in Fig. \ref{IF2}.

In this paper, we develop the phenomenological study of this diffractive $\rho$ + lepton pair production process \footnote{Note that in Ref.~\cite{Pire:2019hos} the process was referred to as diffractive DVCS (DiDVCS).  Here, the name diffractive $\rho$ + lepton pair production is preferred to avoid confusion with the ``plain'' DVCS process.  This also has diffractive contributions at small $x$ where the color dipole fluctuates back into a photon, instead of producing a $\rho$ meson as we consider here.} within the kinematical conditions expected at the future electron-ion colliders. In Sec.~\ref{sec:kin}, we describe the kinematics where we expect our framework to be valid. Sec.~\ref{sec:amp} collects the necessary ingredients for the calculation of the scattering amplitude, and the resulting expressions. Section~\ref{sec:results} presents our results for the cross sections calculations. Finally, Sec.~\ref{sec:conclusion} sums up our conclusions.

\section{Kinematics}
\label{sec:kin}

We study the process (\ref{processGeneral}) 
at large squared photon-nucleon energy $s_{\gamma N} =(q+p_1)^2$, in the forward limit where the $\rho$ meson flies in the same direction as the virtual initial photon and in the kinematical regime of a large rapidity gap between the $\rho^0$ and the final state virtual photon, {\em i.e.} $s_1=(q_\rho+q')^2 \gg s_2= (q'+p_2)^2$. 

  We define 
\begin{align}
&P^\mu = \frac{p_1^\mu + p_2^\mu}{2}\,, &\Delta^\mu = p_2^\mu - p_1^\mu\,,
\end{align}
and decompose momenta on a Sudakov basis  as
\begin{equation}
\label{sudakov1}
v^\mu = \gamma  n^\mu + \delta p^\mu + v_\bot^\mu \,,
\end{equation}
 with $p$ and $n$ the light-cone vectors
\begin{align}
\label{sudakov2}
& p^2=n^2=0\,, &2(pn) \equiv s\,, \nonumber\\
&p^\mu = \frac{\sqrt{s}}{2}(1,0,0,1)\,,  &n^\mu = \frac{\sqrt{s}}{2}(1,0,0,-1) \,,
\end{align}
and
\begin{equation}
\label{sudakov3}
v_\bot^\mu = (0,v^x,v^y,0) \,, \qquad v_\bot^2 = -\vec{v}^2\,.
\end{equation}
The particle momenta read
\begin{eqnarray}
\label{impini}
 q^\mu &=& n^\mu - \frac{Q^2}{s} p^\mu ~, \nonumber \\
 p_1^\mu &=& (1+\xi)\,p^\mu + \frac{M^2}{s(1+\xi)}\,n^\mu~, \nonumber\\
 p_2^\mu &=& (1-\xi)\,p^\mu + \frac{M^2+\vec{\Delta}^2}{s(1-\xi)}n^\mu + \Delta^\mu_\bot ~, \\
 p_\rho^\mu &=& \alpha_\rho \, n^\mu + \frac{m^2_\rho}{\alpha_\rho s}\,p^\mu \,,\nonumber\\
q'^\mu &=& \alpha \, n^\mu + \frac{Q'^2+\vec\Delta^2}{\alpha s}\,p^\mu  -\Delta^\mu_\bot~,\nonumber
\end{eqnarray}
with  
$M$ and $m_\rho$ the masses of the nucleon and of the $\rho$ meson.

The total squared center-of-mass energy of the $\gamma^*$-N system is
\begin{equation}
\label{energysquared}
s_{\gamma N} = (q + p_1)^2 \approx (1+\xi)s - Q^2 + M^2\,.
\end{equation}
Neglecting masses and $\Delta_\perp$, we have
\begin{eqnarray}
\alpha + \alpha_\rho = 1\,.
\end{eqnarray}
Moreover, in the large rapidity gap regime, we have
\begin{align}
&\alpha_\rho \approx  1\,, &\alpha \ll  1 \,.
\end{align}
The squared sub-energies are
\begin{eqnarray}
\label{forwards2}
s_1&=&(p_\rho +q')^2 \approx \frac{Q'^2}{\alpha} \,, \nonumber \\
s_2 &=& (p_2+q')^2 \approx s(1-\xi) \alpha +Q'^2\\
& \approx& s_{\gamma N} \frac{1-\xi}{1+\xi}\alpha + Q'^2\,.\nonumber
\end{eqnarray}
Additionally, momentum conservation in $p^\mu$ leads to
\begin{equation}
    2\xi s = s_1 + Q^2\,,
\end{equation}
and combined with the expressions for $s_1$ and $s_2$ we have
\begin{equation}
    s_{\gamma N} = \frac{s_1s_2}{Q'^2}\,.
\end{equation}
The skewness variable $\xi$ is thus expressed in terms of $Q'^2$ and of the squared sub-energy $s_2$ (we neglect here $Q^2$ compared to $s_{\gamma N} $) as :
\begin{equation}
\label{eq:xi_didvcs}
    \xi \approx \frac{Q'^2}{2s_2-Q'^2}\,,
\end{equation}
which can be compared to a similar expression for $\xi$ in TCS \cite{Berger:2001xd}.

\section{The scattering amplitude}
\label{sec:amp}
\begin{figure} [htb]
    \centering
    \includegraphics[width=.9\textwidth]{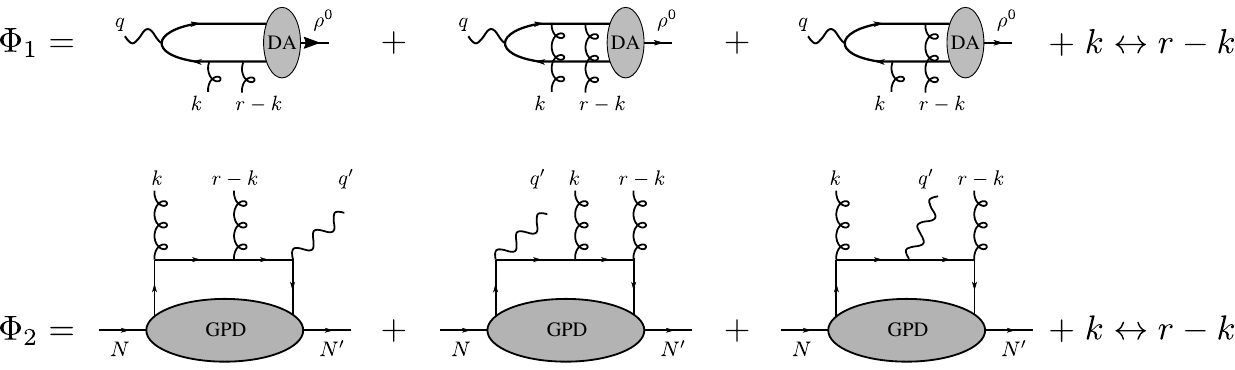} 
    \caption{The leading order Feynman diagrams for the impact factors $\Phi_1$ (upper row) and $\Phi_2$  (bottom row) from Fig.~\ref{fig:process}.  Note the similarities between the hard coefficient function part of the diagrams between both impact factors.  }
    \label{IF2}
\end{figure}
In the kinematical regime described above, it is legitimate to calculate the scattering amplitude in the following way~\cite{Pire:2019hos}. Using the $k_T$ factorization procedure, we write the amplitude as a two-dimensional integral over the transverse components of the exchanged gluon momenta as 
\begin{equation}
\label{ampl}
    \mathcal{M} =\frac{i s}{N_c^2-1}  \int d^2k   \frac{ \Phi_1(k, r-k) \Phi_2(k, r-k)}{(2\pi)^2 k^2(r-k)^2}\,.
\end{equation}
Here the gluon propagator numerators are replaced by $-g^{\mu\nu} \to \frac{-2p^\mu n^\nu}{s}$, $\mu$ (resp. $\nu$) acting on the upper (resp. lower) impact factors. The impact factors  $\Phi_1$ and $\Phi_2$ are calculated (see Fig. \ref{IF2}) within the collinear factorization framework (at $r_\perp=0$) as
\begin{eqnarray}
\label{IF1}
    \Phi_1(k_\perp, -k_\perp) &=& \frac{C_1}{Q} \int  \frac{du  \phi(u) \vec k^2}{\vec k^2+u (1-u) Q^2}\, ,\\
    \Phi_2(k_\perp, -k_\perp) &=& \sum _q\frac{C_2 }{Q'}  \int_{-1}^1 dx ~\theta(\xi^2-x^2)~ 4\xi^2 \vec k^2 \nonumber \\
&&\times \frac{ \bar U(p_2,\lambda_2) [ (n \gamma) e_qH^q(x,\xi,t) +\frac{i \sigma^{n\Delta}}{2M}e_qE^q(x,\xi,t) ]U(p_1,\lambda_1)}{ (x^2-\xi^2) Q'^2 +4 \xi^2 \vec k^2 -i\epsilon } \,, \label{eq:Phi2}
\end{eqnarray}
with   $C_1=2C_F  \sqrt{4\pi\alpha_{\text{em}}} (4\pi \alpha_s) \frac{ f_{\rho^0}}{\sqrt{2}}$ ($f_{\rho^0}=216\,\mbox{MeV}$) and $C_2= - 4C_F (4\pi \alpha_s )\sqrt{4\pi\alpha_{\text{em}}}/s$.
Impact factor $\Phi_1$ has been calculated years ago at leading order \cite{Ginzburg:1985tp, Ivanov:2004pp} and is now known at next to leading order \cite{Ivanov:2005gn}. Its non-perturbative part is the longitudinal $\rho$ meson distribution amplitude (DA) defined, at the leading twist 2, by the matrix element~\cite{Ball:1996tb}
\begin{equation}
\langle 0|\bar{u}(0)\gamma^\mu u(x)|\rho^0(p_\rho,\varepsilon_{\rho L}) \rangle = \frac{1}{\sqrt{2}}p_\rho^\mu f_{\rho^0}\!\!\!\int_0^1 \!\! du\ e^{-iup_\rho\cdot x} \phi(z),
\label{defDArhoL}
\end{equation}
and by a similar expression with opposite sign for d quarks. $\phi(z)$ will be parametrized by either its asymptotical form $\phi_{\text{as}}(z) =6z(1-z)$ or a ``holographic''  $\phi_{\text{h}} = \sqrt{\frac{8}{\pi}} \sqrt {z(1-z)}$ shape~\cite{Forshaw:2012im}. 

Impact factor $\Phi_2$ has been calculated in \cite{Pire:2019hos} as a convolution of the C-odd quark generalized parton distributions (GPDs) and a leading order coefficient function which selects the so-called ERBL (Efremov Radyushkin Brodsky Lepage) region $|x|\leq |\xi|$ in the GPD variables domain of definition \cite{Diehl:2003ny, Belitsky:2005qn}. There is no contribution  from the axial nor from the transversity quark GPDs and no contribution from the gluon GPDs~\cite{Pire:2019hos}.

\section{Cross sections}
\label{sec:results}

\begin{figure} [htb]
    \centering
    \includegraphics[width=.7\textwidth]{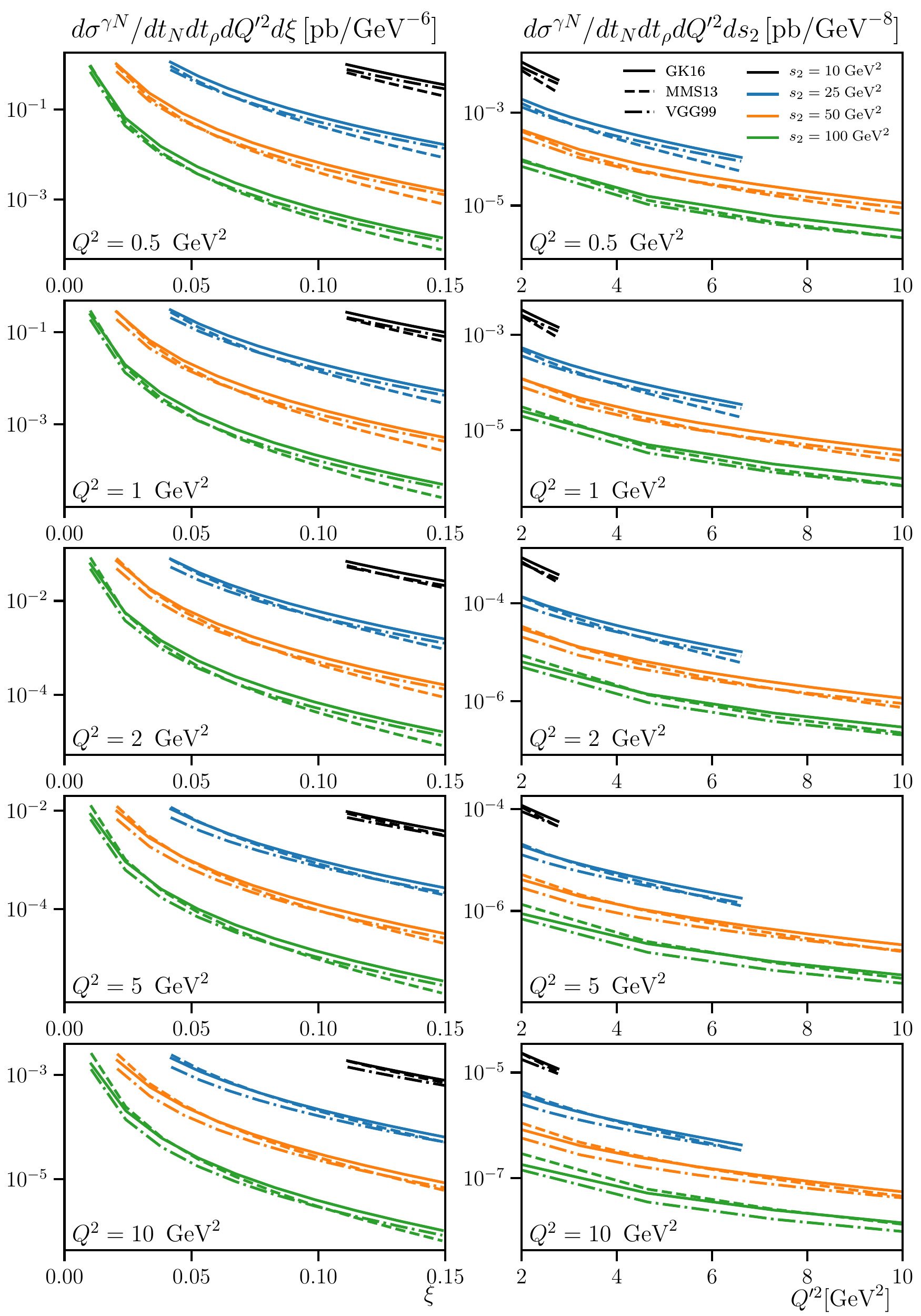}
    \caption{Diffractive $\rho$ + dilepton photoproduction cross sections for different values of $Q^2$ at $t_N=-0.1~\text{GeV}^2$ and $t_\rho=t_{\rho}^{\text{min}}$.  The curves have different values of invariant mass $s_2$ (color) and show three different GPD models (line style), see legend for details.  Left panels show the cross section as a function of $\xi$ (cross section differential in $\xi$), right panels as a function of $Q'^2$ (cross section differential in $s_2$).}
    \label{fig:dsigma_tN0.1}
\end{figure}

\begin{figure} [htb]
    \centering
    \includegraphics[width=.7\textwidth]{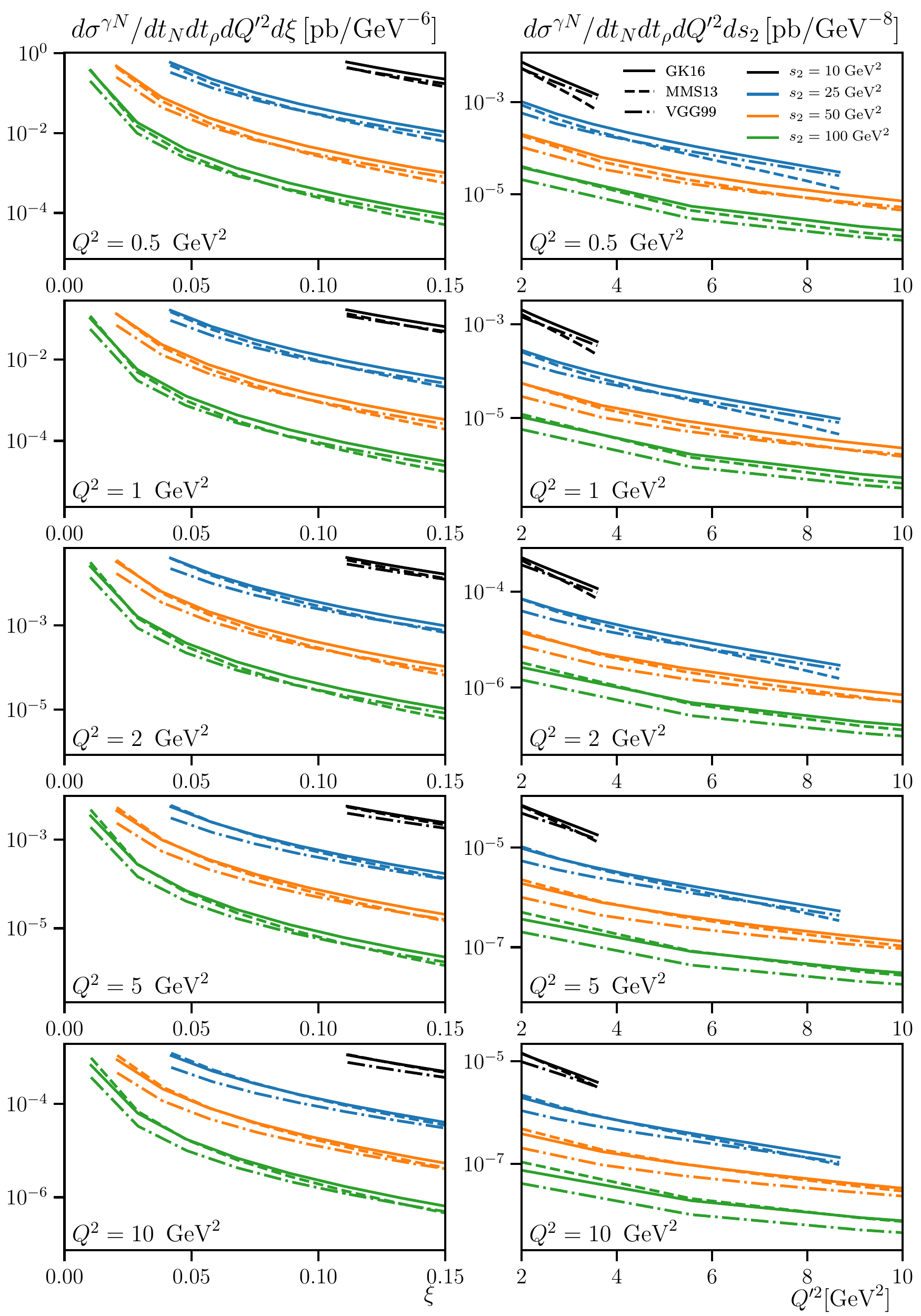}
    \caption{As Fig.~\ref{fig:dsigma_tN0.1} but for $t_N = -0.2~\text{GeV}^2$}
    \label{fig:dsigma_tN0.2}
\end{figure}

\begin{figure} [htb]
    \centering
    \includegraphics[width=.7\textwidth]{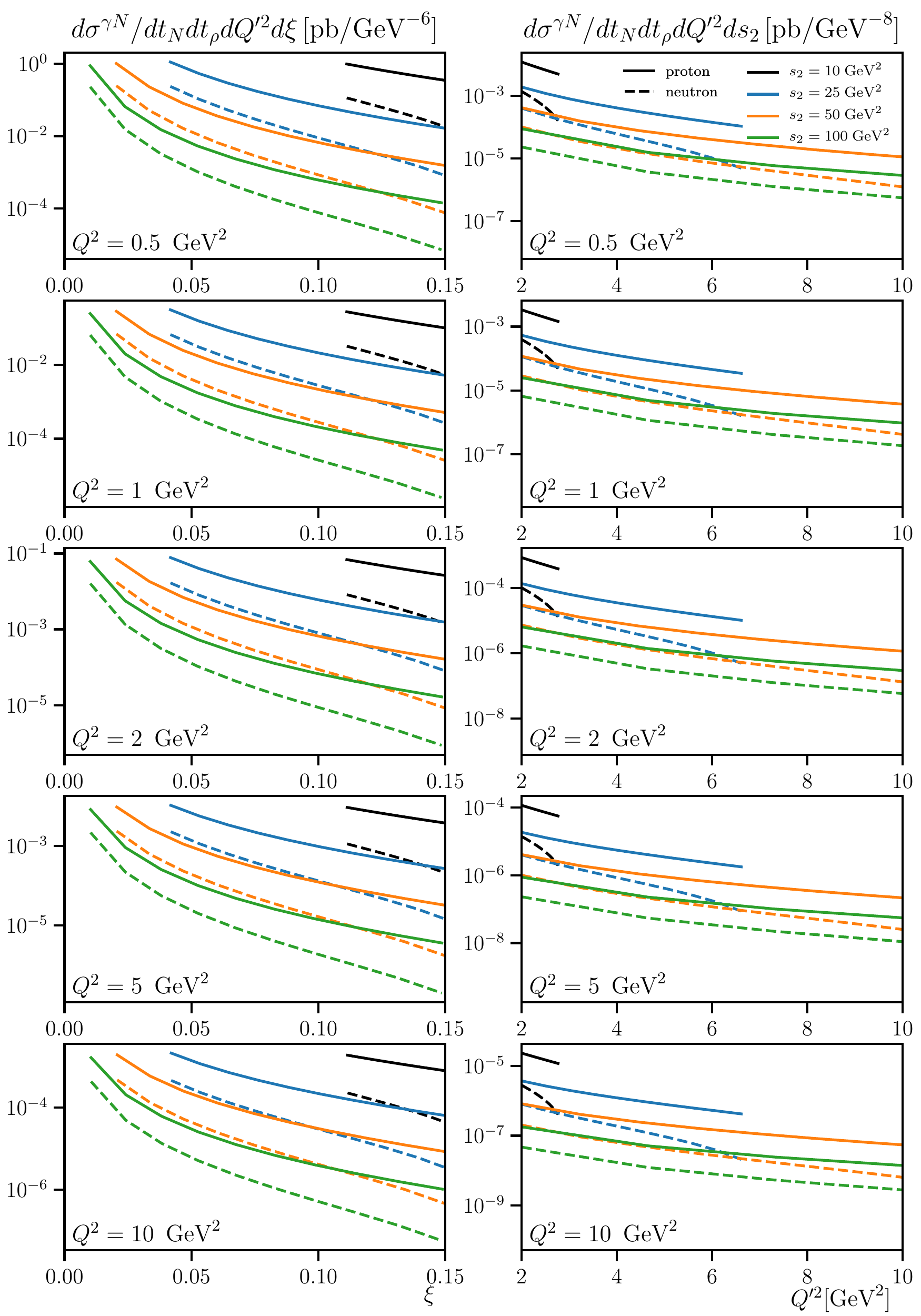}
    \caption{As Fig.~\ref{fig:dsigma_tN0.1} with $t_N=-0.1~\text{GeV}^2$, but comparing free proton and neutron within the GK16 parametrization.}
    \label{fig:dsigma_tN0.1_pn}
\end{figure}

Using the amplitude calculated in Ref.~\cite{Pire:2019hos}, we thus calculate the scattering cross sections for the process (\ref{processGeneral}).
The unpolarized differential cross section for (virtual) photoproduction reads
\begin{align}
\label{eq:xsection_diff_s2}
    d\sigma^{\gamma^* N} &= \frac{1}{256(2\pi)^6 (s_{\gamma N}+Q^2)^2 (s_2-t_\rho)} \overline{\sum}_\text{spins}\; \mathcal{M}^2 \;\; dt_\rho dt_N ds_2 dQ'^2 d\Omega^*_l\nonumber\\
     & = \frac{\alpha_\text{em}}{64(2\pi)^5}\frac{1}{(1+\xi)^2}\frac{1}{Q'^2(s_2-t_\rho)} \left[ (1-\xi)^2|\mathcal{H}(\xi,t)|^2-2\xi^2 \text{Re} \left(\mathcal{H}^{d*}(\xi,t)\mathcal{E}
     ^d(\xi,t)\right) \right. \nonumber\\
     &\left.\qquad -\left( \xi^2+\frac{t_N}{4M^2} \right)|\mathcal{E}^d(\xi,t)|^2\right]\; dt_\rho dt_N ds_2 dQ'^2 d\Omega^*_l
\end{align}
where we summed and averaged over all polarizations, and $\Omega^*_l$ is the solid angle of the final lepton in the lepton pair center-of-mass frame.  Expressions for the Compton form factors (CFFs) $\mathcal{H}^d,\mathcal{E}^d$ are found in Ref.~\cite{Pire:2019hos}. Let us note that a distinctive property of our description of the process under study is the angular distribution of the final state leptons. Since the lepton pair originates from a longitudinally polarized virtual photon, this angular distribution reads, in the center-of-mass system of the lepton pair,
\begin{equation}
    \frac{1}{\sigma} \frac{d\sigma}{d\Omega^*_l} = \frac{3}{8\pi}\sin^2 \theta^*_l\,.
\end{equation}
 This distribution may help to distinguish our process from the production of a misidentified $\pi^+\pi^-$ meson pair.  In the results shown here, we integrate over this lepton solid angle.
 In some panels, we show cross sections differential in $\xi$ instead of $s_2$, using the Jacobian
\begin{equation}
    d\xi = \frac{2\xi^2}{Q'^2}ds_2\,.
\end{equation}
  Note that the $s_{\gamma N}$ dependence in the first equality of Eq.~(\ref{eq:xsection_diff_s2}) cancels with an identical factor in the matrix element~\cite{Pire:2019hos}, which results in a photoproduction cross section that is energy independent, see the second equality of Eq.~(\ref{eq:xsection_diff_s2}).

In the figures below, we show our results for 
the photoproduction cross section of Eq.~(\ref{eq:xsection_diff_s2}) at the minimal value of $-t_\rho$ and at some accessible (small) value of $-t_N$. 
Our collinear framework does not allow to calculate the  $t_\rho$-dependence of the cross section. Phenomenological studies of diffractive electroproduction at HERA have shown \cite{Aaron:2009xp,Chekanov:2007zr} 
that this dependence is very steep and  may be parameterized  as
\begin{equation}
    \frac{1}{\sigma} \frac{d\sigma^{\gamma^* N\to \rho N}}{dt_\rho} \propto \exp[-K(t_\rho^{\text{min}}-t_\rho)] \,,
\end{equation}
with $K \approx 6~ \text{GeV}^{-2} $ for a $\rho N$ final state and $K \approx 2 ~\text{GeV}^{-2}$ for diffractive dissociation, i.e. a $\rho X$ final state. Our case may be seen as intermediate between these two reactions. One may take these two dependences as a range for an educated guess for our case.

Electroproduction cross sections in the approximations used here are given by
\begin{equation}
    d\sigma^{eN} = \frac{\alpha_\text{em}}{2\pi}\frac{y}{Q^2}\frac{\epsilon}{1-\epsilon} d\sigma^{\gamma^* N} dy\,dQ^2\,,
\end{equation}
with only the longitudinal polarization of the incoming virtual photon contributing.

\begin{figure} [htb]
    \centering
    \includegraphics[width=.7\textwidth]{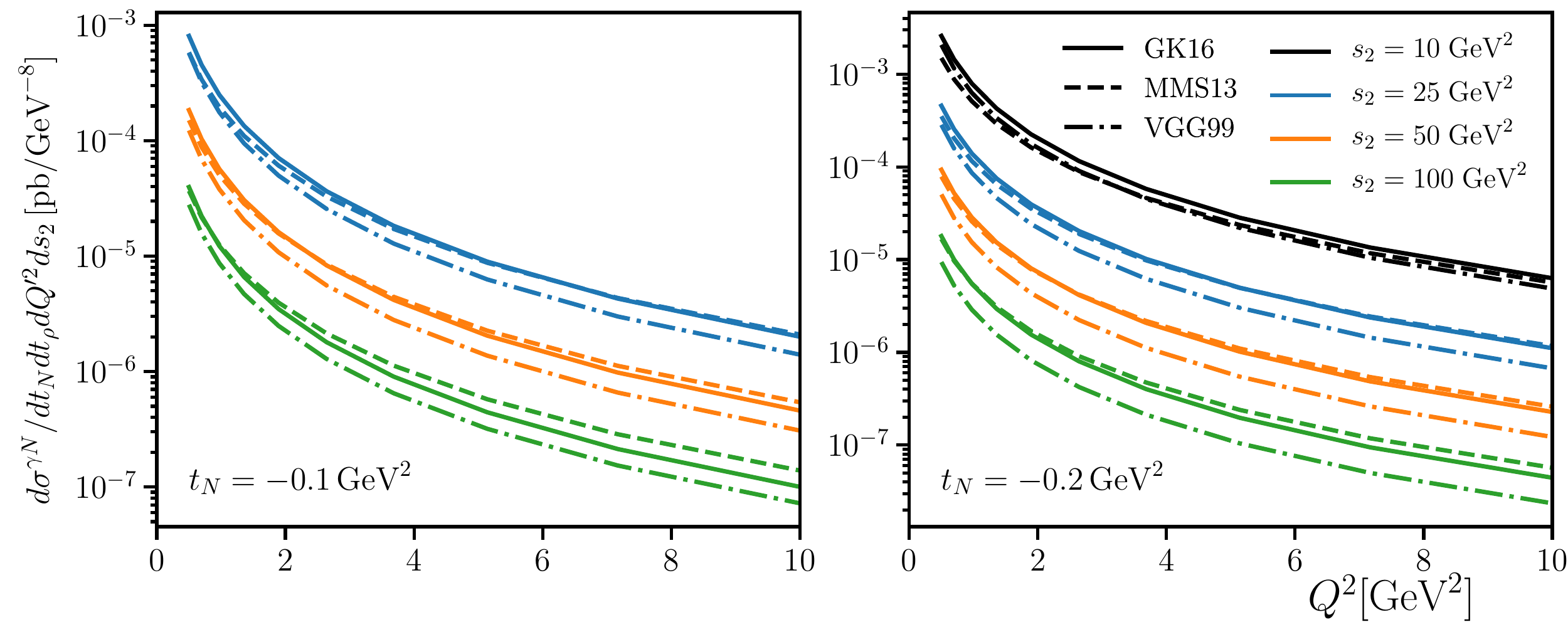}
    \caption{Diffractive $\rho$ + dilepton photoproduction cross sections as a function of $Q^2$ for two values of $t_N$, $Q'^2=3~\text{GeV}^2$ and $t_\rho=t_{\rho}^{\text{min}}$.  Different curves have different values for $s_2$ (color) or GPD models (line style), see legend for details.}
    \label{fig:dsigma_fixQ2in_correct}
\end{figure}

  In Figs.~\ref{fig:dsigma_tN0.1} and \ref{fig:dsigma_tN0.2}, we show photoproduction cross sections where we compare three nucleon GPD parametrizations, namely GK16 \cite{Goloskokov:2007nt,Kroll:2012sm}, VGG \cite{Vanderhaeghen:1999xj} and MMS \cite{Mezrag:2013mya}.  The GPD parametrizations were interfaced through the PARTONS framework~\cite{Berthou:2015oaw}.   Unless mentioned otherwise, all calculations use the asymptotic DA for the $\rho$ vertex.   We show several values of $s_2$, where the minimum $s_2$ value depends on $t_N$ with a larger $|t_N|$ resulting in smaller possible values of $s_2$ (which yield larger cross sections when all other variables are kept fixed).  We show plots of the differential photoproduction cross section Eq.~(\ref{eq:xsection_diff_s2}) as a function of $Q'^2$ or $\xi$. The range in the former maps to a range in the latter for all other kinematic variables fixed, see Eq.~(\ref{eq:xi_didvcs}). We consider a minimum $Q'^2 = 2\,\text{GeV}^2$ to have a sufficient hard scale in the GPD diagram. The maximum $Q'^2$ value is determined by the maximum $\xi$ value allowed by the choice of $t_N$.  In Fig.~\ref{fig:dsigma_tN0.1} we show results for $t_N = -0.1\,\text{GeV}^2$, Fig. \ref{fig:dsigma_tN0.2} has $t_N = -0.2\,\text{GeV}^2$. We show panels for several values of $Q^2$ in each figure, where we consider $Q^2$ values lower than the limit we expect that is required to get a small-sized color dipole in our formalism.  One immediately notices that the cross sections are tiny, even at the kinematics (small $Q^2,Q'^2$; small $s_2$, small $-t_N, -t_\rho$) that maximize its size. The cross section drops quite quickly with $Q'^2$, as expected from Eq.~(\ref{eq:xsection_diff_s2}), with the steepest drop occurring at the lowest $Q'^2$ values.  All three GPD parametrizations result in curves with similar features, with variations in magnitude up to a factor of $\sim 2$, and GK16 exhibiting the slowest drop of the cross sections with $Q'^2$.  In Fig.~\ref{fig:dsigma_tN0.1_pn}, we compare free proton and neutron cross sections using the GK16 parametrization.  Neutron cross sections are significantly smaller than the proton counterparts, and drop off steeper with increasing $\xi$ or $Q'^2$.  In Fig. \ref{fig:dsigma_fixQ2in_correct}, we plot the $Q^2$ dependence at a fixed $Q'^2=3~ \text{GeV}^2$.  We extend the range in $Q^2$ again to values below which our formalism should be valid, and observe that the cross section drops by more than 2 decades over 1 decade of $Q^2$, with the steepest drop at the smallest $Q^2$ values.

\begin{figure} [htb]
    \centering
    \includegraphics[width=.7\textwidth]{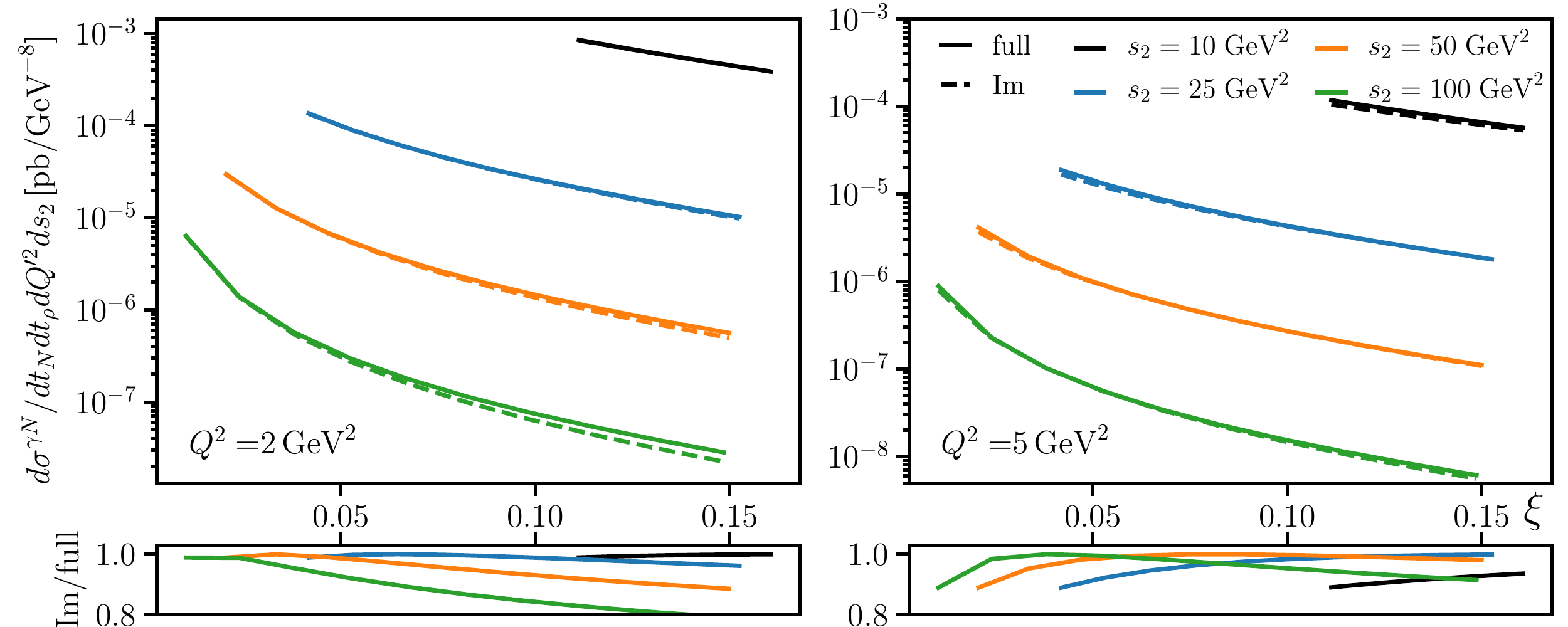}
    \caption{Comparison between a full calculation and one excluding the real part of the CFFs at $t_N=-0.1 \text{GeV}^2$ and $t_\rho=t_{\rho}^{\text{min}}$, using the GK16 GPD parametrization~\cite{Goloskokov:2007nt} at two different $Q^2$ values.  The bottom panels contain the ratio of the cross section calculation with only $\text{Im}({\text{CFFs}})$ to the full calculation. }
    \label{fig:dsigma_nore}
\end{figure}

\begin{figure} [htb]
    \centering
    \includegraphics[width=.7\textwidth]{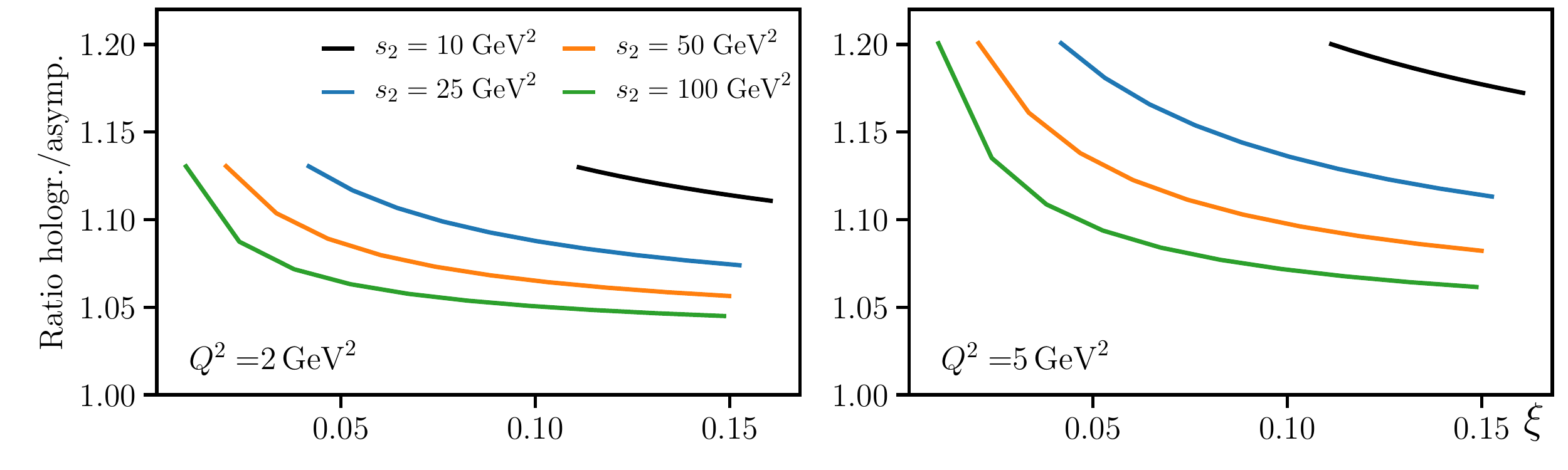}
    \caption{Comparison between two DA parametrizations for the cross sections at the kinematics of Fig.~\ref{fig:dsigma_nore}.  The ratio of the photoproduction cross sections ratios (holographic~\cite{Forshaw:2012im} DA $\propto \sqrt{z\bar{z}}$ and asymptotic DA $\propto z\bar{z}$).  Only the dominant imaginary part of the CFFs was included in this calculation.}
    \label{fig:dsigma_DA}
\end{figure}

\begin{figure} [htb]
    \centering
    \includegraphics[width=.45\textwidth]{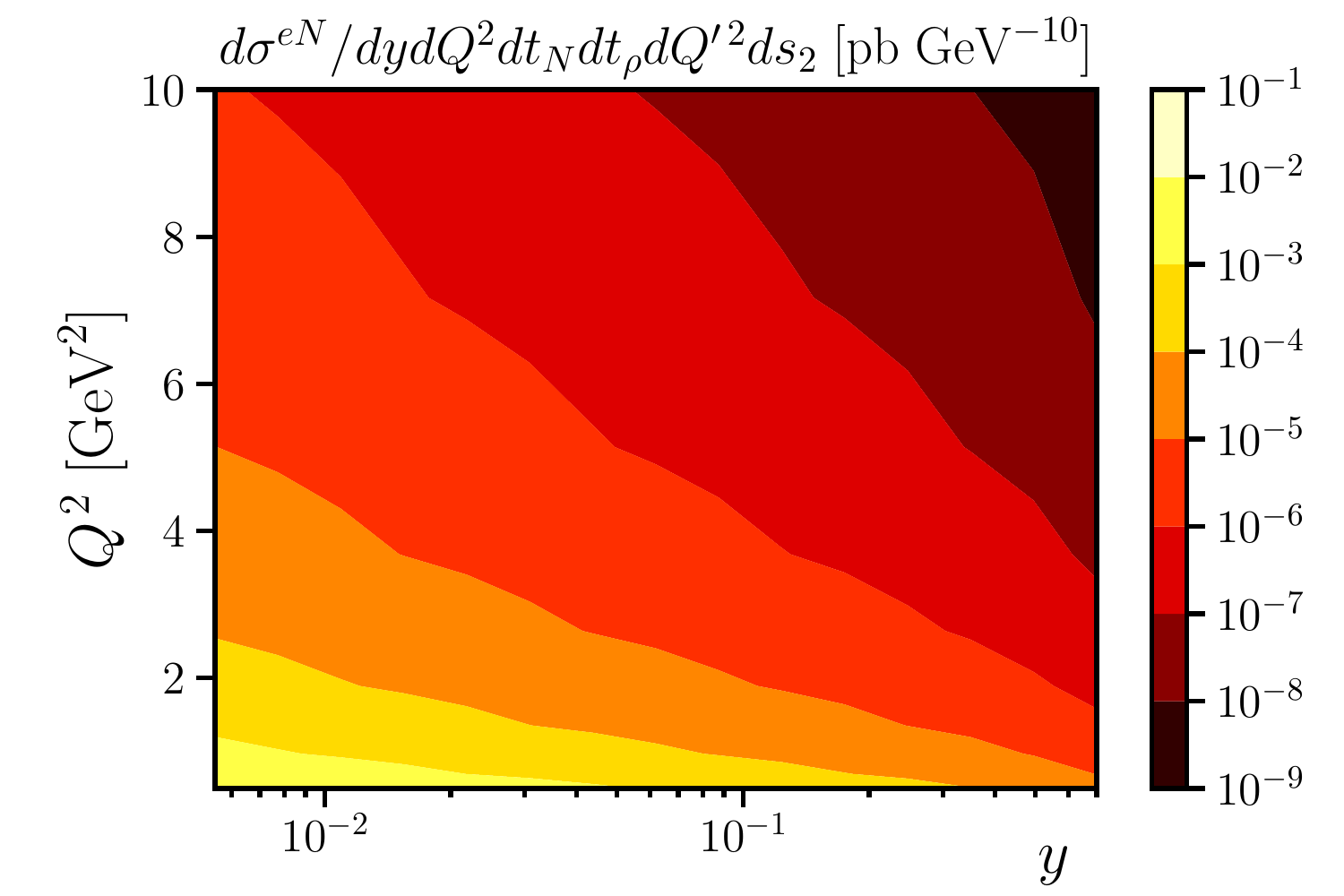}
    \caption{Electroproduction cross section (18 GeV electrons colliding with 275 GeV protons) as a function of $y$ and $Q^{2}$ at $Q^{'\,2}=2~\text{GeV}^2, t_N=-0.1~\text{GeV}^2, t_\rho = t_{\rho}^{\text{min}}, s_2=10~\text{GeV}^2$, using the GK16 GPD parametrization.  }

    \label{fig:electro_Q2out}
\end{figure}

\begin{figure} [htb]
    \centering
    \includegraphics[width=.45\textwidth]{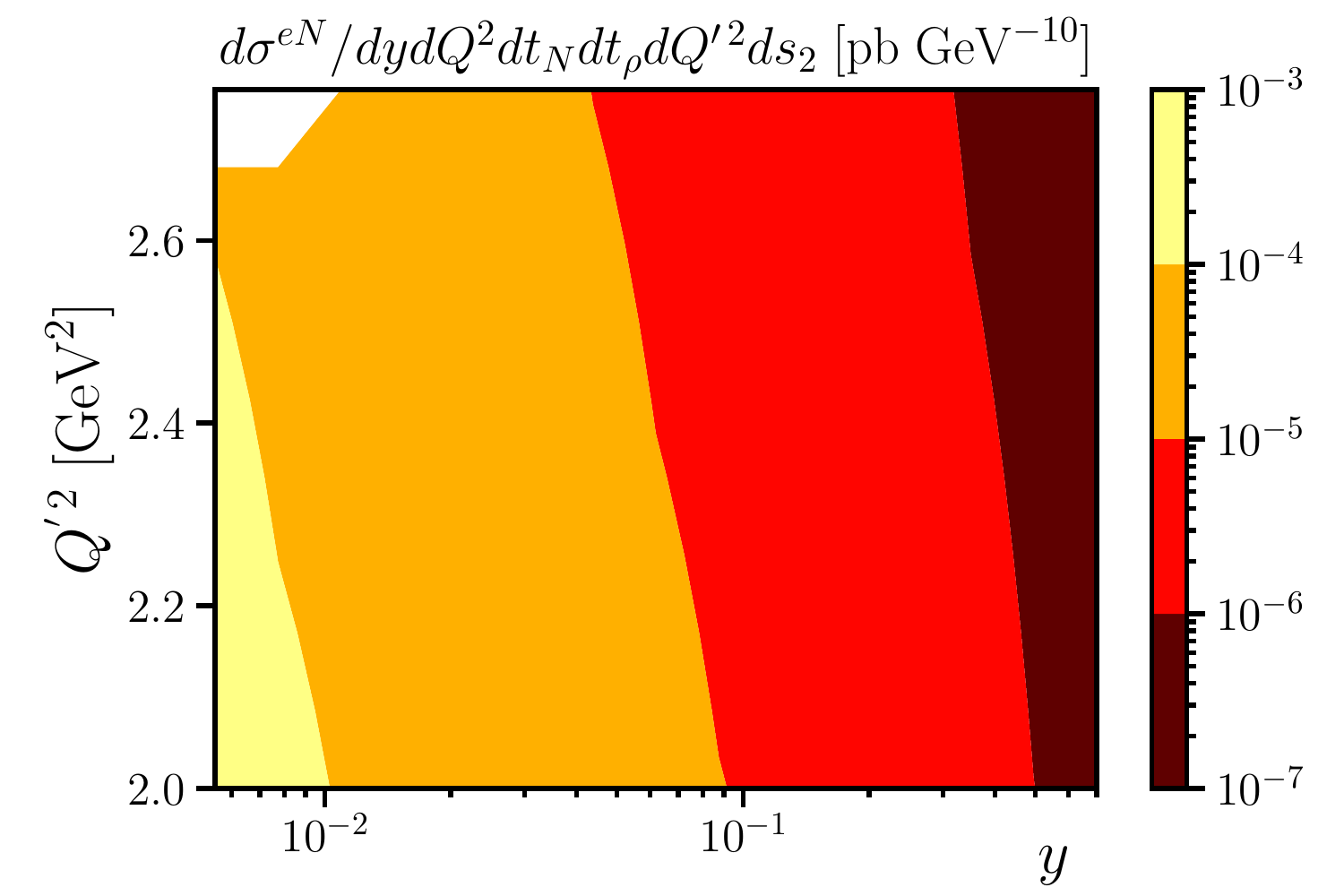}
    \caption{Electroproduction cross section (18 GeV electrons colliding with 275 GeV protons) as a function of $y$ and $Q^{'\,2}$ at $Q^2=2~\text{GeV}^2, t_N=-0.1~\text{GeV}^2, t_\rho = t_{\rho}^{\text{min}}, s_2=10~\text{GeV}^2$ using the GK16 GPD parametrization. The white area corresponds to unphysical kinematics with $\xi>\xi_\text{max}$.}
    \label{fig:electro}
\end{figure}

  Figure~\ref{fig:dsigma_nore} illustrates that the imaginary part of the Compton form factors -- originating from the imaginary part of Eq.~(\ref{eq:Phi2})-- entering in the amplitude dominate the cross section, contributing 80\% or more of the total strength.  Two choices of DA parametrization (asymptotic and holographic) are compared in Fig.~\ref{fig:dsigma_DA}, with the difference in results between the two  20\% or smaller, far smaller than the difference between the different GPD parametrizations.  Electroproduction cross sections for the high-energy configuration of the planned U.S. electron-ion collider at Brookhaven National Laboratory, are shown in Fig.~\ref{fig:electro_Q2out} as a function of $y$ and $Q^2$ and in Fig.~\ref{fig:electro} as a function of $y$ and $Q'^2$.  We observe that minimizing any of these variables maximizes the cross sections but that rates are still small, showing limited promise at the moment of this process being worthy of detailed studies at an electron-ion collider.  Note, however, the remarks with regard to NLO corrections in the concluding section.

\section{Conclusion}
\label{sec:conclusion}
Our studies show that the cross sections at leading order for the diffractive $\rho$ + dilepton production process are quite small, which can make a straightforward analysis of the process at the luminosities of planned electron-ion collider facilities very difficult.  In terms of non-perturbative inputs, the calculations show much greater sensitivity to the nucleon GPD input than to the $\rho$ meson DA one.  This GPD model sensitivity is due to the quite unique fact (see however \cite{Ivanov:2002jj,Enberg:2006he,Cosyn:2020kfe} for a similar dependence) that the amplitude only depends on their behaviour in the ERBL region, which is quite unrestricted by current data analysis of the DVCS process.  The cross section is dominated by the imaginary part of the Compton form factors and is maximized at small values of the hard scales $Q^2, Q'^2$ where higher order corrections to the formalism would be needed.
The small magnitude of our cross sections deters us from studying in detail competing processes such as a quasi-Bethe-Heitler contribution where the dilepton originates from a virtual photon radiated from the electron line. Our kinematics do not favor such a production process, since a lepton pair emitted from the electron line is likely to lead to a large value of $s_2 \ge s_{\gamma N}$ which is quite opposite to the kinematical domain here explored. This quasi-Bethe-Heitler process is thus expected to be significantly suppressed compared to the QCD contribution, in contradistinction to the double DVCS process \cite{Belitsky:2003fj}.

The fact that  the cross section is rather small must be blamed on the presence of two hard scales ($Q^2$ and $Q'^2$) which each play a crucial role in keeping a part of the process to be controlled by small-size hadronic configurations. This is to be contrasted with the diffractive process \cite{Ivanov:2002jj,Cosyn:2020kfe} $e N \to e' \rho \pi N'$ with a large transverse momentum $\rho$ meson, where this large transverse momentum was the single large scale controlling the perturbative treatment of both impact factors. The size of the diffractive $\rho$ + dilepton production cross section is more reminiscent of the  deep electroproduction of a large invariant mass  dilepton \cite{Guidal:2002kt,Belitsky:2002tf} or diphoton \cite{Pedrak:2020mfm}.
The expected cross sections increase rather quickly at small $Q^2$, which opens the need for a better understanding of the diffractive $\rho$ + dilepton reaction in the region where the Pomeron becomes soft while the large mass of the final state lepton pair still pleads for a collinear factorization approach of the lowest impact factor which probes the nucleon GPDs. We shall tackle this problem in future studies.

The study of NLO QCD effects, both for the Pomeron exchange propagators using the BFKL framework and the impact factors in the collinear framework, remains to be done. This is not an easy task although complete NLO studies for quite similar reactions already exist \cite{Ivanov:2004pp,Enberg:2005eq,Ivanov:2005gn}, but with the important difference of the {\em timelike} vs. {\em spacelike} nature of the virtual photon, where one expects analytic continuation effects to be quite important~\cite{Pire:2011st, Muller:2012yq}. Although one cannot accurately predict the order of magnitude of this correction without completing the calculation, it is quite fair to say that such NLO corrections should not give more than a 100\% correction to the Born order calculation of the amplitude without demanding a resummation procedure to be carried out before a theoretical estimate can reliably be quoted. This would undoubtedly be a very interesting but quite intricate problem. Measuring an anomalously large experimental rate for our process would indeed be valuable information to question the validity of the theoretical approach developed here, namely the hybrid framework where part of the amplitude is discussed in terms of $k_T$-factorization while the other part relies on collinear QCD factorization.

\acknowledgments 

\noindent
We thank Janus Chwastowski, Maxime Defurne, Rafal Staszewski and Lech Szymanowski  for   useful discussions.

\bibliography{references}

\end{document}